\documentclass[11pt]{article}
\usepackage[utf8]{inputenc}
\usepackage{authblk}
\usepackage{setspace}
\usepackage[margin=1.25in]{geometry}
\usepackage{graphicx}
\usepackage{subcaption}
\usepackage{amsmath}
\usepackage{physics}

\usepackage{braket}
\usepackage[font=footnotesize,labelfont=bf]{caption}

\usepackage[
  style=chem-rsc, 
  citestyle=numeric-comp,
  sorting=none
  ]{biblatex}
\title{
\textbf{The symmetric quasi-classical model using on-the-fly time-dependent density functional theory within the Tamm-Dancoff approximation}} 
\author[1*]{\textbf{Justin J. Talbot}}
\author[1,2]{\textbf{Martin Head-Gordon}}
\author[1]{\textbf{Stephen J. Cotton}}

\affil[1]{Department of Chemistry, University of California, Berkeley, California 94720, United States}
\affil[2]{Chemical Sciences Division, Lawrence Berkeley National Laboratory, Berkeley, California, 94720, United States }
\affil[*]{\textbf{Email: justin.talbot@berkeley.edu}}
 \usepackage[labelfont=bf]{caption} 

\date{}

\begin{document}
\maketitle

\begin{abstract}
The primary computational challenge when simulating nonadiabatic \textit{ab initio} molecular dynamics is the unfavorable compute costs of electronic structure calculations with molecular size. Simple electronic structure theories, like time-dependent density functional theory within the Tamm-Dancoff approximation (TDDFT/TDA), alleviate this cost for moderately sized molecular systems simulated on realistic time scales. Although TDDFT/TDA does have some limitations in accuracy, an appealing feature is that, in addition to including electron correlation through the use of a density functional, the cost of calculating analytic nuclear gradients and nonadiabatic coupling vectors is often computationally feasible even for moderately-sized basis sets. In this work, some of the benefits and limitations of TDDFT/TDA are discussed and analyzed with regard to its applicability as a ``back-end'' electronic structure method for the symmetric quasi-classical Meyer-Miller model (SQC/MM). In order to investigate the benefits and limitations of TDDFT/TDA, SQC/MM is employed to predict and analyze a prototypical example of excited-state hydrogen transfer in gas-phase malonaldehyde. Then, the ring-opening dynamics of selenophene are simulated which highlight some of the deficiencies of TDDFT/TDA. Additionally, some new algorithms are proposed that speed up the calculation of analytic nuclear gradients and nonadiabatic coupling vectors for a set of excited electronic states.
\end{abstract}

\section{Introduction}

A detailed, molecular-level description of nonadiabaticity at the \textit{ab initio} level is useful when elucidating many important photoinduced chemical and physical processes \cite{curchod2018ab,tapavicza2013ab,schuurman2018dynamics,levine2007isomerization,domcke2011conical,jasper2004introductory,matsika2011nonadiabatic}. Simulating complex nonadiabatic dynamics processes however, is oftentimes prohibited by the computational cost of electronic structure calculations which can have high polynomial or exponential scalings with system size---particularly if high-order excitations and/or electron correlation is required for an accurate description of the excited electronic state. A simple approach that seeks to alleviate this cost is to represent each excited electronic state using only single excitations in the configuration interaction wavefunction while including the effects of electron correlation through the use of a Kohn-Sham (KS) reference determinant. This approach, known as time-dependent density functional theory within the Tamm-Dancoff approximation (TDDFT/TDA), has improved scalings with system size, e.g. $\sim\mathcal{O}(N^2-N^3)$ per state, compared to many other correlated methods and meaningful predictions of nonadiabatic processes that both explain and predict experimental observables are possible \cite{hirata1999time}. 

An appealing feature of TDDFT/TDA from a dynamics perspective is the efficiency when computing analytic nuclear gradients and nonadiabatic couplings which are used to construct nuclear forces and first-order derivative coupling vectors. \cite{maurice1998single,liu2010parallel,fatehi2011analytic,zhang2014analytic,ou2015derivative}. Conveniently, in the Tamm-Dancoff approximation, the machinery for calculating analytic first-order derivative coupling vectors is exactly the same as the analytic nuclear gradient \cite{zhang2014analytic}. Often this results in the same computational routines being used for both calculations. The main drawback when using analytic gradient routines however, is the leading quadratic computational cost that grows with the number of excited electronic states. While some newly-proposed methods have been proposed that avoid this approach altogether for first-order derivative couplings, using overlap based \cite{richter2011sharc,plasser2016efficient} and finite difference approximations \cite{ryabinkin2015fast,pittner2009optimization}, calculating these couplings analytically clearly offers the greatest accuracy. With algorithmic improvements this cost is tractable even for moderately-sized molecular systems that require propagating trajectories on multiple excited electronic states.

While many methodologies have been proposed that propagate the equations of motion for the electronic and nuclear degrees of freedom (DOF) \cite{tully1990molecular,wang2016recent,kapral2016surface,li2005ab,ding2015ab,saita2012fly,isborn2007time,curchod2016communication,mignolet2018walk}, with varying degrees of complexity \cite{tully2012perspective,crespo2018recent}, a computationally efficient and in many cases sufficiently accurate approach is to propagate both DOF on an equal footing using classical Hamiltonian mechanics \cite{meyer1979classical}. One such approach is the symmetric quasi-classical Meyer-Miller model (SQC/MM) which quantizes the electronic degrees of freedom in the Meyer-Miller (MM) Hamiltonian using a set of predefined windowing functions which are applied, symmetrically, both to sample initial conditions, and to estimate electronic state populations (and/or coherences) at prescribed times during the classical vibronic dynamics evolution \cite{cotton2013symmetrical2,cotton2013symmetrical}. While typically the SQC/MM approach has been used to model the complex nonadiabatic dynamics of model systems in the diabatic representation, recent years have seen significant progress in the development of the SQC/MM model to predict the dynamics of general molecular systems in the adiabatic representation. This has been through improved adiabatic EOM \cite{cotton2017adiabatic} and also some initial realistic calculations employing ``on-the-fly'' electronic structure theories \cite{hu2021fly,weight2021ab,zhou2019quasi,talbot2022dynamic}. 

In this work, Meyer-Miller dynamics, as employed in both the standard Ehrenfest method as well as the SQC model, are implemented and analyzed using ``on-the-fly'' electronic structure theory in the Q-Chem software package \cite{epifanovsky2021software}. The implementation of this methodology required algorithmic improvements that reduce the cost when evaluating analytic nuclear gradients and first-order derivative coupling vectors for multiple electronic states. Using SQC/MM with TDDFT/TDA, a simple analysis of the implemented algorithms is performed by simulating the population dynamics and geometric rearrangements that mediate excited-state hydrogen transfer in malonaldehyde. Then, as a more complex illustration of this approach, SQC/MM is employed to make predictions of the excited-state ring-opening dynamics of selenophene upon photoexcitation which highlights some limitations of TDDFT/TDA when modeling bond breaking. 

\section{Methods}

The following notation is used throughout this work: $I$ and $J$ denote adiabatic Born-Oppenheimer electronic states where an electron has been excited from $i$,$j$,$k$,$\ldots$ occupied KS orbitals to $a$,$b$,$c$,$\ldots$ virtual orbitals in the reference determinant. $\mu$,$\nu$,$\lambda$, $\sigma$,$\ldots$ are indices denoting atomic orbital (AO) basis functions. $\hat{A}^{[\textbf{R}]}$ denotes the full Cartesian derivative of the operator $\hat{A}$ with respect to Cartesian nuclear DOF $\textbf{R}$ which also indicates differentiation of the KS orbital coefficients. All electronic states, orbitals, and basis functions are assumed to be real unless otherwise noted. 

\subsection{SQC/MM Nonadiabatic Dynamics}

The classical Meyer-Miller Hamiltonian maps the electronic DOF in a nonadiabatically-coupled dynamic system to a collection of classical harmonic oscillators. The SQC/MM approach combines this mapping with a simple, yet effective quantization protocol for the electronic DOF along a classical trajectory. When electronic structure calculations are used for the nuclear forces and couplings, the adiabatic basis is most amenable. The MM Hamiltonian expressed in this basis is
\providecommand{\mbf}[1]{\mathbf{#1}}
\providecommand{\bP}{\mbf P}
\providecommand{\bR}{\mbf R}
\providecommand{\hf}{\tfrac 12}
\providecommand{\bmu}{\boldsymbol{\mu}}
\providecommand{\Eq}[1]{Eq.~\ref{e:#1}}
\providecommand{\Eqs}[2]{Eqs.~\ref{e:#1} and~\ref{e:#2}}
\providecommand{\Eqsthree}[3]{Eqs.~\ref{e:#1}, \ref{e:#2}, and~\ref{e:#3}}
\begin{equation}\label{e:MMa}
    \textbf{H}(\textbf{x},\textbf{p},\textbf{R},\textbf{P}) = \frac{1}{2\bmu} \left(\textbf{P} + \Delta\textbf{P}\right)^2 + V_{\text{eff}}(\textbf{x},\textbf{p},\textbf{R}),
\end{equation}
\noindent where $\mbf{R,\, P}$ denote the positions and momenta of the $3N$-Cartesian nuclear DOF with atomic masses $\bmu$. In the MM framework, the nuclei move on an effective potential energy surface given by
\begin{equation} \label{e:Veff}
    V_{\text{eff}} (\textbf{x},\textbf{p},\textbf{R}) = \sum_I^F \bigg( \hf p_I^2 + \hf x_I^2 - \gamma_I \bigg)E_I(\textbf{R}), 
\end{equation}
\noindent where $\{x_I,p_I\}$  are the positions and momenta of the ``electronic oscillators'' defining a set of $F$ adiabatic electronic states each with energy $E_I$. $\{\gamma_I\}$ denotes a set of zero point energy (ZPE) parameters in the electronic DOF. In the adiabatic representation, the nuclear momentum $\bP$ arises in combination with a nonadiabatic coupling vector potential 
\begin{equation*} 
    \Delta \textbf{P} (\textbf{x},\textbf{p},\textbf{R}) = \sum_{I<J}^F (x_Ip_J - x_Jp_I)\; \textbf{d}_{IJ}(\textbf{R}) ,
\end{equation*}
which depends explicitly on the standard first-order derivative coupling vector $\textbf{d}_{IJ}(\textbf{R}) = \left\langle{\Psi_I}|\vec{\nabla}_\textbf{R}\Psi_J\right\rangle$ between adiabatic electronic states $\Psi_I$ and $\Psi_J$. The occupation-weighted effective potential shown in \Eq{Veff} is commonly symmetrized 
\begin{equation}\label{e:Veffs}
   V_{\text{eff}}(\textbf{x},\textbf{p},\textbf{R}) = \frac{1}{F}\sum_I^{F}E_I(\textbf{R})
    + \frac{1}{F}\sum_{I<J}^F 
    \left(p_I^2 - p_J^2 + x_I^2 - x_J^2\right)\, \left(E_I(\textbf{R})-E_J(\textbf{R})\right) ,
\end{equation}
which sets the energy zero and guarantees the electronic dynamics are independent of energy scale. 

The canonical equations of motion (EOM) are obtained by applying Hamilton's equations 
\begin{equation}\label{e:Heq}
    \dot{x}_I = \frac{\partial \textbf{H}}{\partial p_I}, \quad \dot{p}_I = -\frac{\partial \textbf{H}}{\partial x_I}, \quad \dot{\textbf{R}} = \frac{\partial \textbf{H}}{\partial \textbf{P}}, \quad \dot{\textbf{P}} = -\frac{\partial \textbf{H}}{\partial \textbf{R}}
\end{equation}
\noindent to the adiabatic MM Hamiltonian in \Eq{MMa} producing dynamically-consistent, canonical coordinates and momenta in both the nuclear and electronic DOF. An apparent drawback of using the adiabatic basis however, is that Hamilton's equations introduce \emph{second}-derivative nonadiabatic coupling \emph{matrices} into the EOM. As recently shown however \cite{cotton2017adiabatic}, the explicit calculation of these second-derivative nonadiabatic coupling matrices can be avoided entirely by employing a simple change of variables from the canonical nuclear momentum to the so-called ``kinematic'' nuclear momentum \\
\begin{equation*}
    \textbf{P}_{\textbf{kin}} = \textbf{P} + \Delta\textbf{P}.
\end{equation*}
\newcommand{\Pkin}{\mathbf P_{\text{kin}}}%
Although $\Pkin$ is not canonically-conjugate to $\mathbf R$ it can be utilized in generating exactly the same Hamiltonian dynamics via the following kinematic EOM:
\begin{subequations}\label{e:knEOM}
\begin{align}
    \dot{x}_I &= \frac{p_I}{F}\sum_J^{F}E_I(\textbf{R}) - E_J(\textbf{R}) + x_J\textbf{d}_{JI}(\textbf{R}) \cdot \frac{\textbf{P}_{\text{kin}}}{\mu}, \\
    \dot{p}_I &= -\frac{x_I}{F}\sum_J^{F}E_I(\textbf{R}) - E_J(\textbf{R}) + p_J\textbf{d}_{JI} (\textbf{R}) \cdot \frac{\textbf{P}_{\text{kin}}}{\mu}, \\
    \dot{\textbf{R}} &=  \frac{\textbf{P}_{\text{kin}}}{\mu}, 
\end{align}
\begin{equation}
        \dot{\textbf{P}}_{\text{kin}} = -\frac{\partial V_{\text{eff}}}{\partial \textbf{R}} - \sum_{IJ}\left(\frac{1}{2}p_Ip_J + \frac{1}{2}x_Ix_J\right)(E_J(\textbf{R}) - E_I(\textbf{R}))\textbf{d}_{IJ}(\textbf{R}).
\end{equation}
\end{subequations}
These kinematic EOM, advantageously, contain only the first-order derivative couplings $\mbf d_{JI}(\textbf{R})$, but are nevertheless exactly equivalent to the EOM obtained after employing \Eq{Heq} which includes both first- and second-order couplings. 

In the SQC/MM approach, quantization of the classical Hamiltonian dynamics produced by \Eq{knEOM} is done symmetrically, i.e., with respect to both the initial and final values of the dynamical electronic variables. Quantization is accomplished, initially by Monte Carlo sampling actions from a ``windowing'' function defined by the SQC model. The quantization at the prescribed final times is accomplished by ``binning'' the final time-evolved actions according to the windowing function. In Q-Chem, the triangle windowing model \cite{cotton2016new} is available with the option to use a $\gamma$-adjustment procedure, exactly as described in Ref. \cite{cotton2019trajectory}, except that here the $\gamma$-adjustment procedure is employed with the kinematic EOM of \Eq{knEOM}. The key point of the $\gamma$-adjustment procedure is to set the $\{\gamma_I\}$ in \Eq{Veff} per DOF (and per trajectory), so that the initial forces on the nuclei are that of the initial pure quantum state---i.e., the single-surface forces. Ehrenfest simulations are also available where the dynamics of these are equivalent to the SQC calculations, but instead of using symmetric windowing functions for selecting initial conditions and estimating final populations, the Ehrenfest method uses integer initial electronic action variables with $\gamma = 0$ and uses the \emph{values} of these action variables at each desired final time to estimate the electronic state populations instead of evaluating whether the actions fall within a window function.

The nuclear EOM (\Eq{knEOM}c and \Eq{knEOM}d) are integrated numerically using a traditional velocity-Verlet integrator. The electronic EOM (shown in \Eq{knEOM}a and \Eq{knEOM}b) are integrated using a semi-analytic scheme that solves the time-dependent electronic Schr{\"o}dinger equation at each time step with the nuclear coordinates and momenta as momentarily fixed. This is equivalent to solving the following set of first-order coupled differential equations
\begin{equation}\label{e:TDSE}
    i\dot{\textbf{C}} = \textbf{H}\textbf{C}
\end{equation}
\noindent where $\textbf{C}$ are the set of time-dependent electronic amplitudes 
\begin{equation}\label{e:elecosc}
    C_I(t) = \frac{1}{\sqrt{2}}\bigg(x_I(t) + ip_I(t)\bigg),
\end{equation}
\noindent that are defined according to the electronic oscillator variables, and $\textbf{H}$ is the electronic Hamiltonian with matrix elements
\begin{equation*}
    H_{IJ} = \frac{1}{F}\sum_K^{F}\bigg(E_I(\textbf{R}) - E_K(\textbf{R})\bigg)\delta_{IJ} - i \textbf{d}_{JI} (\textbf{R}) \cdot \frac{\textbf{P}_{\text{kin}}}{\mu},
\end{equation*}
\noindent expressed in the adiabatic basis. The time-dependent electronic amplitudes are obtained by diagonalizing $\textbf{H}$, at each time step, and writing the solution as a complex exponential
\begin{equation}\label{e:TDSEsol}
    \textbf{C}_{t+1} = \textbf{U}e^{-i\boldsymbol{\epsilon}\Delta t}\textbf{U}^{\dag}\textbf{C}_{t},
\end{equation}
\noindent where $\Delta t$ denotes the time step, $\textbf{U}$ are the eigenvectors, and $\epsilon$ are the eigenvalues of \textbf{H}. The real and imaginary components of $\textbf{C}$ constitute the time-stepped electronic oscillator coordinates and momenta (scaled by $1 /\sqrt{2}$), respectively.

The time-stepped solution to the time-dependent Schr{\"o}dinger equation shown in \Eq{TDSEsol} is exact for fixed nuclei, however construction of the Hamiltonian matrix assumes that the nuclei are fixed during the electronic update. This is an approximation but seems to have a negligible impact on the accuracy of the electronic dynamics and allows propagation of both the electronic and nuclear DOF with the same time step. For problematic situations, i.e. when $\textbf{H}$ changes rapidly in time, a higher-order numerical integrator may be required at the additional expense of introducing a shorter electronic time step. 

\subsection{Analytic Gradients and Nonadiabatic Couplings}

Time propagation of the electronic and nuclear DOF requires nuclear gradients and first-order derivative coupling vectors for a set of adiabatic electronic states. The first-order derivative coupling vector between states $\Psi_I$ and $\Psi_J$ is calculated using the Hellmann-Feynman theorem \cite{domcke2011conical} 
\begin{equation*}
    \textbf{d}_{IJ}(\textbf{R})  =  \frac{\textbf{h}_{IJ}}{\omega_J-\omega_I},
\end{equation*}
\noindent where $\omega_I$ and $\omega_J$ are TDDFT/TDA excitation energies and
\begin{equation*}
\textbf h_{IJ} \equiv \bra{\Psi_I}\hat{H}^{[\textbf{R}]}\ket{\Psi_J}
\end{equation*}
\noindent is the nonadiabatic coupling vector. In the TDDFT/TDA formalism, the excited state wavefunction is a projection of the eigenfunctions of the electronic Hamiltonian $\hat{H}$ onto the space of single excitations
\begin{equation*} 
    \ket{\Psi_I} = \sum_{ia}X_{ai}^{I}\ket{\Phi_i^a},
\end{equation*}
\noindent where $\ket{\Phi_i^a}$ denotes a singly-excited determinant after promoting an electron from an occupied orbital $i$ to a virtual orbital $a$ in the KS reference. The excitation amplitudes $X_{ai}^I = \left\langle{\Psi_I}|\Phi_i^a\right\rangle$ are obtained by solving the following eigenvalue equation
\begin{equation*}
    \textbf{A}\textbf{X}^{I} = \omega_I\textbf{X}^{I},
\end{equation*}
\noindent where \textbf{A} is a single excitation Hamiltonian which is Hermitian in the Tamm-Dancoff approximation since the corresponding excitation and de-excitation amplitudes have been uncoupled.

An appealing property of TDDFT/TDA is that the analytic expression for the nonadiabatic coupling is similar to the excited state analytic gradient \cite{zhang2014analytic}
\begin{equation}\label{e:hIJexp} 
    \textbf{h}_{IJ} = \sum_{ijab} X_{ai}^IA_{ai,bj}^{[\textbf{R}]}X_{bj}^J = \sum_{ijab} X_{ai}^I \bigg[F_{ab}^{[\textbf{R}]}\delta_{ij}-F_{ij}^{[\textbf{R}]}\delta_{ab} + \Pi_{ia,bj}^{[\textbf{R}]} + \Omega_{ai,bj}^{[\textbf{R}]}\bigg]X_{bj}^J,
\end{equation}
\noindent where $\boldsymbol{F}$ is the KS Fock matrix, $\boldsymbol{\Pi}$ is the two-electron integral tensor, and $\boldsymbol{\Omega}$ denotes the response of the exchange-correlation Fock matrix after a perturbation in the one-particle density matrix \cite{hirata1999time,liu2010parallel}. \Eq{hIJexp} is a generalized Hellmann-Feynman-type expression which one might assume is \emph{not} valid because the wavefunctions employed are not eigenfunctions of the electronic Hamiltonian; however, it has been shown in Ref. \cite{fatehi2012derivative} that the additional non-Hellmann-Feynman terms that arise after projecting the eigenfunctions onto the space of single excitations renders the first-order derivative coupling dependent on overall translational motion which is obviously unphysical. The procedure advised in Ref. \cite{zhang2014analytic} is to simply leave these additional non-Hellmann-Feynman terms out of the expression for the nonadiabatic coupling which is justified by the introduction of electronic translation factors into the electronic EOM \cite{fatehi2011analytic}.

Evaluating analytic nuclear gradients and nonadiabatic coupling vectors requires building one- and two-particle density matrices. Constructing the required density matrices allows the nonadiabatic coupling in \Eq{hIJexp} to be expressed in a compact form
\begin{equation}\label{e:hij}
    \textbf{h}_{IJ} = \textbf{P}_{\Delta}^{IJ'} \cdot \textbf{H}^{[\textbf{R}]} + \boldsymbol{\Gamma}^{IJ'} \cdot \boldsymbol{\Pi}^{[\textbf{R}]} + \textbf{W}^{IJ'} \cdot \textbf{S}^{[\textbf{R}]} \\ + \textbf{P}_{\Delta}^{IJ'} \cdot \boldsymbol{F_{xc}^{[\textbf{R}]}} + \textbf{T}^{I\dag} \cdot \boldsymbol{\Omega}^{[\textbf{R}]} \cdot \textbf{T}^{J},
\end{equation}
\noindent where $\textbf{H}^{[\textbf{R}]}$,  $\textbf{S}^{[\textbf{R}]}$, and $\boldsymbol{F_{xc}^{[\textbf{R}]}}$ denotes the Cartesian derivatives of the core Hamiltonian, overlap, and exchange-correlation Fock integrals and $\boldsymbol{\Pi}^{[\textbf{R}]}$ and $\boldsymbol{\Omega}^{[\textbf{R}]}$ denotes the Cartesian two-electron and exchange-correlation response integral derivatives, respectively. Expressions for the required density matrices and further derivations of the components of $\textbf{h}_{IJ}$ are provided in Appendix~A.

In a nonadiabatic dynamics simulation, evaluating \Eq{hij} for multiple electronic states at each time step can quickly become the dominant computational expense. Q-Chem already contains efficient analytic gradient and nonadiabatic coupling routines that evaluate \textbf{h}$_{IJ}$ between any single \emph{pair} of states \cite{liu2010parallel,zhang2014analytic}. One approach, in a multi-state protocol, would be to simply use this code to evaluate \textbf{h}$_{IJ}$ between all combinations of pairs of the electronic states during a trajectory. Such an approach (referred to as scheme~I) involves re-calculating all integrals and integral derivatives for each density matrix which is clearly not ideal as this amounts to the most computationally expensive step being needlessly repeated for each pair. An improved approach would be to simply build all of the required density matrices up front and contract them all simultaneously, thereby \emph{re-using} already computed integrals and integral derivatives. We have implemented this approach, referred to as scheme~II, because it leads to significant cost improvements as the most computationally expensive step (i.e. evaluating integrals and integral derivatives) is performed once for a \emph{common set} of density matrices. 

\begin{figure}[t!]
  \centering
  \includegraphics[height=7.5cm]{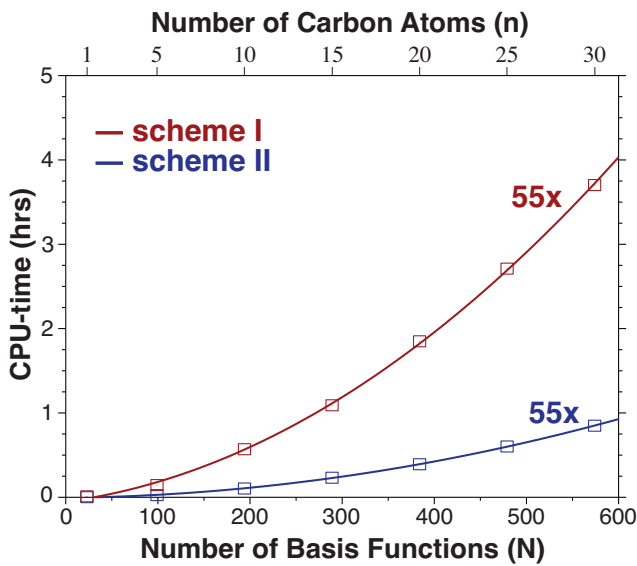}
  \caption{Timing benchmarks as a function of basis set (\textbf{N}) and increasing chain length (\textbf{n}) for a series of alkane molecules. The timings reported are the full evaluation of the set of analytic nuclear gradient and first-order derivative coupling vectors ($55$ in total). The ten lowest-energy singlet excited states of each system were included. The red curves (scheme~I) and the blue curves (scheme~II) correspond to a polynomial least-squares fit.}
  \label{Fig1}
\end{figure}

Timing results that illustrate the speedups possible when employing scheme~II are shown in Fig.~\ref{Fig1}. The molecular system used for the timing analysis was a series of alkane molecules where the length was systematically increased. \Eq{hij} was used to compute the analytic nuclear gradients and nonadiabatic coupling vectors for the ten lowest-energy singlet electronic states in each system (corresponding to $55$ vectors in total). The PBE density functional was employed for all calculations with a 6-31G$^*$ basis set. All timings were benchmarked on a single thread/core $3.6$ GHz Intel Core i9 processor. As evident from Fig.~\ref{Fig1}, the savings after employing scheme~II is significant particularly when more than $300$ basis functions are used (for $300$ basis functions scheme~II takes approximately $15$ minutes while scheme~I takes approximately $1$ hour). Fitting the timings to a quadratic polynomial reveals that, in general this factor of approximately $4\times$ speedup remains even when $600$ basis functions are used. Further analysis of the fit is provided in Appendix~B. 

\subsection{State-Following and Conical Intersections}

A precondition for electronic transitions between adiabatic electronic states is a non-vanishing first-order derivative coupling. However, approaching symmetry-allowed conical intersections and un-avoided crossings during a dynamics trajectory are also possible which can present a bookkeeping challenge when tracking and identifying adiabatic states. Therefore, a protocol is required to ensure that the electronic wavefunction doesn't instantaneously change character during a trajectory by allowing the system to correctly pass through these allowed degeneracies. Furthermore, a protocol is required that ensures the phase of the electronic wavefunction is consistent throughout the trajectory, which, in turn, ensures that the first-order derivative couplings are smooth functions of the nuclear DOF.

A simple approach, that is independent of the phase of the KS orbitals, is to assign electronic states based on the difference between their attachment and detachment density matrices at subsequent time steps \cite{closser2014simulations}. For the multi-state tracking protocol employed here, an approximate overlap matrix is constructed from the similarity metric
\begin{equation*}
        M_{IJ}= 
\begin{cases}
     1 - \left\|\begin{matrix}\Delta \textbf{A}^{IJ}\end{matrix}\right\|,& \text{if } \left\|\begin{matrix}\Delta \textbf{D}^{IJ}\end{matrix}\right\| \leq \left\|\begin{matrix}\Delta \textbf{A}^{IJ}\end{matrix}\right\|\\[10pt]
     1 - \left\|\begin{matrix}\Delta \textbf{D}^{IJ}\end{matrix}\right\|,              & \text{otherwise},
\end{cases}
\end{equation*}
where $\Delta \textbf{A}^{IJ} = \textbf{A}_{t+1}^{I} - \textbf{A}_{t}^{J}$ is the difference between the attachment density for state $I$ at time step $t+1$ and the attachment density for state $J$ at time step $t$. Similarly, $\Delta \textbf{D}^{IJ} = \textbf{D}_{t+1}^{I} - \textbf{D}_{t}^{J}$ is the difference between the detachment density for state $I$ at time step $t+1$ and the detachment density for state $J$ at time step $t$ and $\left\|\begin{matrix} \cdot \end{matrix}\right\|$ denotes the spectral norm.

The matrix element $M_{IJ}$ is an approximate electronic state overlap and in most cases is sufficient for state tracking. However, as a result of this matrix being constructed from differing electronic basis sets at consecutive time steps, this matrix should be projected onto a common basis. This is accomplished by taking the singular value decomposition of $M_{IJ}$
\begin{equation*}
    \textbf{M} = \textbf{U} \cdot \boldsymbol{\Sigma} \cdot \textbf{V}^{T},
\end{equation*}
where $\textbf{U}$ are the left singular vectors which map the electronic basis at $t+1$ onto a common basis, $\boldsymbol{\Sigma}$ are the singular values of $\textbf{M}$, and $\textbf{V}^{T}$ are the right singular vectors which map the electronic basis at time $t$ onto the common basis. With the singular vectors in hand, constructing the nearest orthogonal matrix representation \cite{zhanserkeev2021adiabatic,zhang2015nearest} to this approximate overlap matrix
\begin{equation*} 
    \textbf{Q} = \textbf{U} \cdot \textbf{V}^{T},
\end{equation*}
defines an orthogonalized similarity metric. Assigning state character to specific adiabatic states is done with a ``Min-Cost'' assignment algorithm that permutes the elements of $\textbf{Q}$ until the trace is maximized  \cite{song2020first,carpaneto1988algorithms}. Once the trace is maximized, the energies and corresponding amplitudes are swapped according to the unique set of indices that resulted in the maximum trace. A consistent overall phase for the amplitudes is enforced directly from the overlap, at time $t$ and at $t+1$, between the transition density matrices (see \Eq{Td}).

\section{Model Systems}

The simulations of nonadiabatically-mediated molecular rearrangements presented here (proton transfer and ring-opening) are intended as illustrative examples of modeling small molecular systems with the methodologies presented in this work. The modeling of the treated relaxation pathways is rigorous and accurate at the level of theory presented here but there are some relevant relaxation pathways that, for various reasons, have not been included in these simulations. For example, one such pathway for both malonaldehyde and selenophene is nonradiative decay from the optically-dark $S_1$ electronic state to the ground state which is known to occur on time scales greater than $50$ fs \cite{coe2006ab,pederzoli2017new}. Nonradiative decay pathways to the ground state have not been included here because, in such cases, the ground electronic state is multi-reference and TDDFT/TDA is known to incorrectly predict topologies of the resulting conical intersections \cite{herbert2016beyond}. Likewise, the simulations presented here do not include spin-orbit coupling which is known to be physically relevant in both systems \cite{list2020probing,pederzoli2017new}. Nevertheless, these examples constitute important demonstrative examples of the SQC/MM methodology and the new implementation in Q-Chem.

\subsection{Excited-State Hydrogen Transfer in Malonaldehyde}

Malonaldehyde is a simple prototypical example of excited-state hydrogen transfer with many theoretical studies analyzing and identifying the complex interconversion and intersystem crossing pathways \cite{schroder2014calculation,nandipati2019controlled}. Geometrically, malonaldehyde favors a closed ring structure where an intramolecular hydrogen bond is formed between neighboring carbonyl groups. While a substantial barrier for hydrogen transfer is evident on the $S_0$ and the optically-forbidden $S_1(n\pi^*)$ and $S_3(n\pi^*)$ potential energy surfaces, hydrogen transfer on the optically-bright $S_2(\pi\pi^*)$ state is believed to be barrier-less where the bonding hydrogen favors an equidistant configuration between the two oxygen terminals \cite{sobolewski1999photophysics,coe2006ab}.

Simulating the ultrafast interconversion efficiency after photoexcitation to the $S_2$ state, i.e. for $t< 50$ fs, is well suited for TDDFT/TDA since conical intersections with the ground and triplet electronic states aren't yet accessible and the population transfers quite rapidly to the $S_1$ state. In the longer-time regime (i.e. $t > 50$ fs), alternate pathways to the ground and low-lying triplet states emerge after substantial population has transferred into the $S_1$ state. Identifying the structural rearrangements necessary to activate these relaxation pathways have led to some debate including a proposed three-state conical intersection \cite{coe2007ab,coe2005competitive,coe2006ab}. Recently, List et al. combined both experiments with theory to identify and assign these relaxation pathways using molecular dynamics and x-ray absorption measurements \cite{list2020probing}.

In the current work, a treatment of malonaldehyde's short-time $S_2$ relaxation pathway is presented based on Ehrenfest, SQC/MM, and fewest-switches surface hoping (FSSH) trajectories in the gas phase, with a particular focus on illustrating the mechanics of the SQC/MM methodology and new Q-Chem implementation. For all simulations, trajectories are initialized by sampling nuclear positions and momenta directly from a 0K ground-state harmonic oscillator Wigner distribution, with the electronic degrees of freedom initialized as described above for the Meyer-Miller methods. The electronic degrees of freedom for the FSSH trajectories were initialized on $S_2$ with integer actions in exactly the same way as the Ehrenfest trajectories. The reduced masses and harmonic frequencies used to construct the Wigner distribution were calculated from the minimum energy geometry on the ground electronic state potential energy surface. The PBE0 density functional was employed with the 6-31G$^*$ basis set. The Wigner sampled positions and momenta were allowed to propagate via Meyer-Miller and FSSH dynamics on the three coupled potential energy surfaces: $S_3$, $S_2$, and $S_1$ with a $0.24$ fs time step for $t = 60$ fs. The gamma-adjustment protocol was employed for the SQC/MM trajectories. 

\begin{figure}[t!]
\centering
  \includegraphics[height=8.5cm]{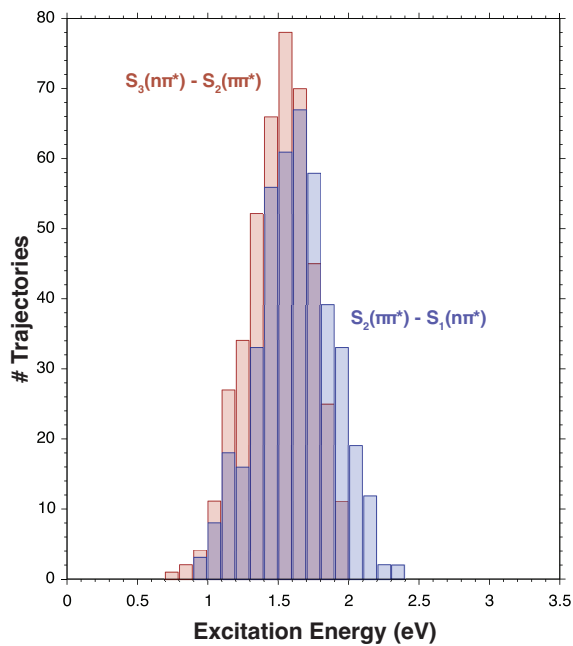}
  \caption{The binned energy differences between the $S_2(\pi\pi^*)$ and $S_1(n\pi^*)$ excitation energies (blue) and the $S_3(n\pi^*)$ and $S_2(\pi\pi^*)$ excitation energies (red) in malonaldehyde. Approximately $400$ initial positions were sampled from a 0K ground-state harmonic oscillator Wigner distribution.
  }
  \label{Fig2}
\end{figure}

Mapping the nonadiabatic dynamics at each time step onto the adiabatic basis states was distinct for malonaldehyde since the initially populated $S_2$ state, assuming Franck-Condon vertical excitation after sampling the Wigner distribution, is energetically well separated from both the $S_1$ and $S_3$ states. To illustrate this, Fig. \ref{Fig2} shows the energy differences in the Franck-Condon region between the $S_2$-$S_1$ and $S_3$-$S_2$ electronic states which are on average $\approx1.5$ eV suggesting that the initially populated $S_2$ state is constructed of mainly $\pi\pi^*$ character.  

\begin{figure*}
\centering
  \includegraphics[height=8.5cm]{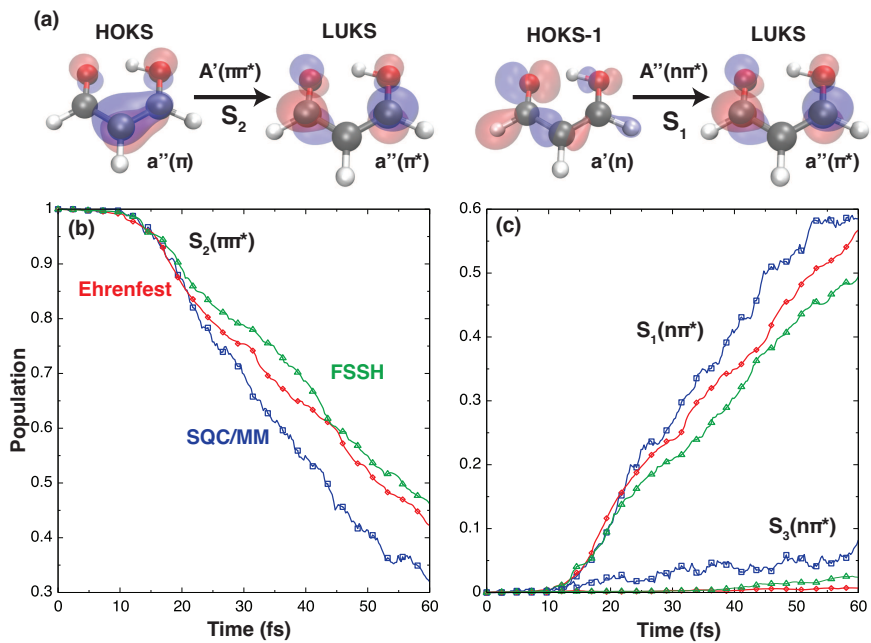}
  \caption{\textbf{(a)} The top contributing KS orbital excitations of the $S_2 (\pi\pi^*)$ and $S_1 (n\pi^*)$ electronic states of malonaldehyde. The population dynamics of the $S_2 (\pi\pi^*)$ electronic state \textbf{(b)} and the $S_1 (n\pi^*)$ and $S_3(n\pi^*)$ electronic states \textbf{(c)} simulated with the Ehrenfest (red, $\approx 200$ trajectories), SQC/MM (blue, $\approx 400$ trajectories), and FSSH (green, $\approx 200$ trajectories) methods.
}
  \label{Fig3}
\end{figure*}

The primary orbital contributions for the two most active states ($S_2$ and $S_1$) are shown in Fig.~\ref{Fig3}(a). The highest-occupied KS orbital (HOKS) is comprised of out-of-plane $\pi$-type orbitals on the acceptor and donor oxygen atoms with a $\pi$ bonding orbital on the carbon backbone. The lowest-unoccupied KS orbital (LUKS) combines similar out-of-plane $\pi$-type orbitals on the acceptor and donor oxygen atoms with a $\pi$ anti-bonding orbital on the carbon backbone. The HOKS-1 orbital is an anti-bonding $\sigma$-type orbitals on the oxygen atoms. The $S_2$ and $S_1$ states are comprised primarily of excitations from the HOKS and HOKS-1 orbitals to the LUKS, respectively. 

The population dynamics are shown in Fig.~\ref{Fig3}(b) and Fig.~\ref{Fig3}(c). All three methods, Ehrenfest, SQC/MM, and FSSH predict a similar decay out of $S_2$ with SQC/MM predicting slightly more population transfer to the $S_1$ state compared with the FSSH and Ehrenfest predictions. Ehrenfest predicts a similar population transfer to SQC/MM up to $t=20$ fs then the most significant deviations between all three methods occurs between $t=20$ and $t=40$ fs. After $t=40$ fs, Ehrenfest predictions of the population transfer out of $S_2$ are closer to the FSSH results. Over the course of the trajectories, less than $10$\% of the population transfers to the $S_3$ state with SQC/MM predicting slightly more population transfer than Ehrenfest or FSSH. 

\begin{figure}[ht]
\centering
  \includegraphics[height=11cm]{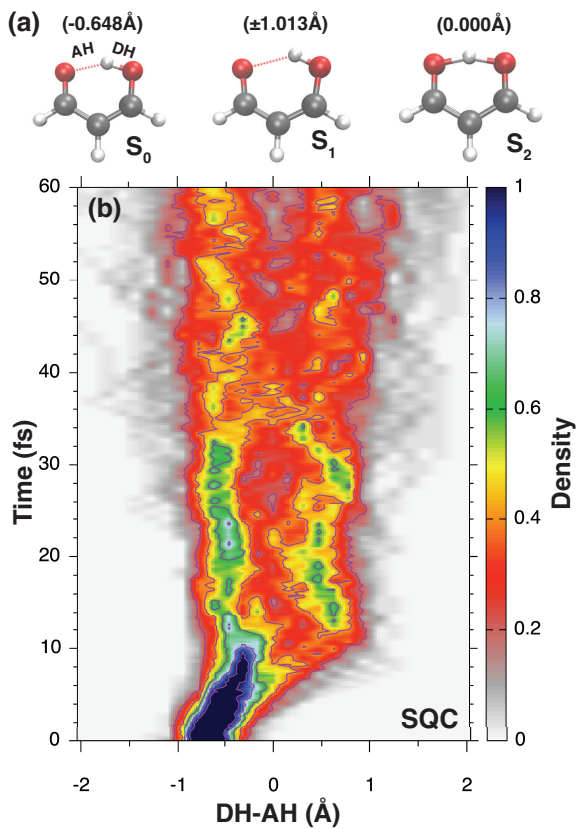}
  \caption{\textbf{(a)} Stationary points on the ground, first-, and second-excited state potential energy surfaces of malonaldehyde. The minimum difference between the donor-hydrogen (DH) and acceptor-hydrogen (AH) bond lengths at each stationary point and for each electronic state is shown in parenthesis. \textbf{(b)} The time dependence of the difference (DH-AH) bond lengths monitored during the SQC/MM trajectories. 
  }
  \label{Fig4}
\end{figure}

For reference and comparison with Ref. \cite{coe2006ab}, the donor minus acceptor hydrogen bond length (DH-AH) was calculated at stationary points on the potential energy surfaces and monitored during the MM dynamics, i.e. those initialized with the SQC procedure (see Fig.~\ref{Fig4}(a)). On the $S_0$ potential energy surface the minimized (DH-AH) bond length is $-0.648$~{\AA}. The DH-AH distance is significantly lengthened, suggesting localization on one of the terminals, on the $S_1$ potential energy surface ($\pm 1.013$~{\AA}). On the $S_2$ potential energy surface, the hydrogen is equidistant between the two oxygen atoms and is free to shuffle between donor and acceptor. The degree of hydrogen transfer during the dynamics is substantial, as shown in Fig.~\ref{Fig4}(b), where the hydrogen shuffles back and forth rapidly from $t=10$ to $t=20$ fs. Once substantial population has transferred into the $S_1$ state (at $t>20$ fs), the hydrogen atom begins localization on either of the oxygen terminals as evidenced by density depletion near DH-AH$ = 0$~{\AA}. 

The simulations of the $S_2$ relaxation pathways in malonaldehyde presented here should serve as a guide when using the Ehrenfest or SQC/MM methods in Q-Chem. An interesting result, in addition to the significant and rapid population transfer that occurs from the $S_2$ to the $S_1$ state, is the dispersion of DH-AH bond lengths throughout the SQC/MM simulations. Since an effective potential forms between $t=40$ and $t=60$ fs, i.e. a weighted average of $S_2$ and $S_1$ with significant $S_2$ character, the difference bond length rarely reaches the optimized value of DH-AH$ = \pm 1.013$ {\AA} on the $S_1$ surface and and is more probable between $\pm 0.75$ {\AA}. This is not surprising considering that by $t=60$ fs the occupation-weighted potential has approximately $60\%$ $S_1$ character (as shown in Fig. \ref{Fig3}(b)) and there is sufficient DH-AH density near approximately $60\%$ of the optimal value.  

\subsection{Ring-Opening Dynamics of Selenophene}

Heterocyclic compounds are important building blocks for many modern technologies, from biomedical applications \cite{chandrashekarachar2017impotrtance,alcolea2016chalcogen} to electronic devices \cite{feng2022recent,park2014effects,son2008analyzing}; and various properties of these compounds can be explored in the gas phase where a detailed, atomistic treatment is feasible with quasi-classical molecular dynamics methods \cite{bhattacherjee2018photoinduced,jankowska2021ultrafast,barbatti2008nonadiabatic,meng2018uv,xie2017position}. Typically, these species exhibit optically-bright $\pi\pi^*$ states which are short lived and involve a competing series of internal conversion pathways to nearby $\pi\pi^*$ and $\pi\sigma^*$ states that promote both ring-puckered and ring-opened configurations, respectively. Additional competing pathways emerge in ring-opened configurations, i.e. after sufficient energy has transferred into $\pi\sigma^*$ configurations, as these systems are known to undergo intersystem crossing to nearby triplet states and nonradiative decay to the ground electronic state \cite{ashfold2017exploring,borissov2021recent}. Of the heterocyclic compounds, five-membered chalcogen containing ring systems have been extensively studied using nonadiabatic dynamics methods and, in such systems, this series of competing pathways between ring-opened and ring-puckered configurations is particularly evident \cite{schalk2018ring,li2019computational,gromov2011ab,gromov2010theoretical,schalk2020competition,gavrilov2008deactivation,stenrup2011computational,prlj2015excited,schnappinger2017ab,weinkauf2008ultrafast}. 

\begin{figure*}
\centering
  \includegraphics[height=7cm]{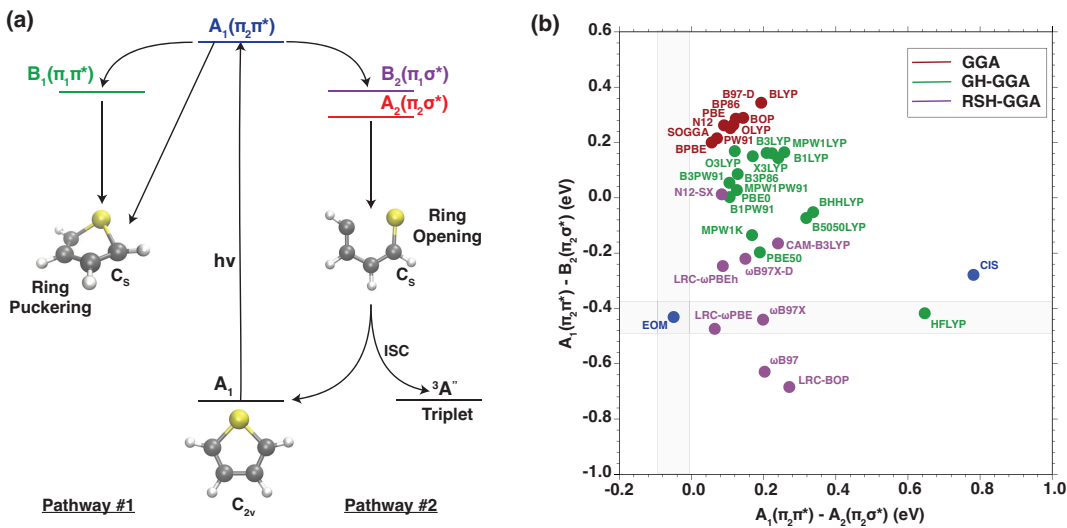}
  \caption{\textbf{(a)} A simplified schematic of the available nonadiabatic pathways to ring opening and ring puckering in selenophene. \textbf{(b)} Energy differences across between the $A_1(\pi_2\pi^*)$, $A_2(\pi_2\sigma^*)$, and $B_2(\pi_1\sigma^*)$ excitation energies referenced from the $C_{2v}$ geometry on the ground electronic state for each method. The EOM-EE-CCSD results are shown in blue for reference.
}
  \label{Fig5}
\end{figure*}

Selenophene (in the gas phase) provides an illustrative example of these types of competing electronically nonadiabatic dynamics. A simplified schematic, after photoexcitation to the optically-bright singlet $A_1(\pi_2\pi^*)$ electronic state, is shown in Fig.~\ref{Fig5}(a). The first excitation pathway consists of either staying on the $A_1$ state or undergoing internal conversion to the singlet $B_1(\pi_1\pi^*)$ state resulting in a distortion of the planar geometry and ring puckering. The second pathway consists of undergoing internal conversion to either the singlet $B_2(\pi_1\sigma^*)$ or the singlet $A_2(\pi_2\sigma^*)$ state. Once sufficient population has transferred into one of these $\pi\sigma^*$ states, ring opening can occur. In ring opened configurations, additional pathways emerge which result in either ring closing after decay back to the singlet ground electronic state or intersystem crossing to low-lying triplet states. 

In order to elucidate the competing pathways in selenophene, excitation energy differences were calculated across the standard hierarchy of density functionals with each energy difference referenced from the optimized $C_{2v}$ geometry with the 6-311G$^{**}$ basis set (see Fig.~\ref{Fig5}(b)). For comparison, the results from the EOM-EE-CCSD/aug-cc-pVTZ level are shown in blue. Using the EOM-EE-CCSD differences as a benchmark, only range-separated density functionals (RSH-GGA) give comparable results where the closest energy differences are predicted by the LRC-$\omega$PBE and $\omega$B97X functionals. Generalized gradient approximations and their global hybrid variants (GGA and GH-GGA) systematically overestimate the energy differences in comparison. Since the LRC-$\omega$PBE/6-311G$^{**}$ level has the closest energy difference when compared to the benchmark, this functional and basis set was chosen for all simulations. 

\begin{figure*}
\centering
  \includegraphics[height=8.5cm]{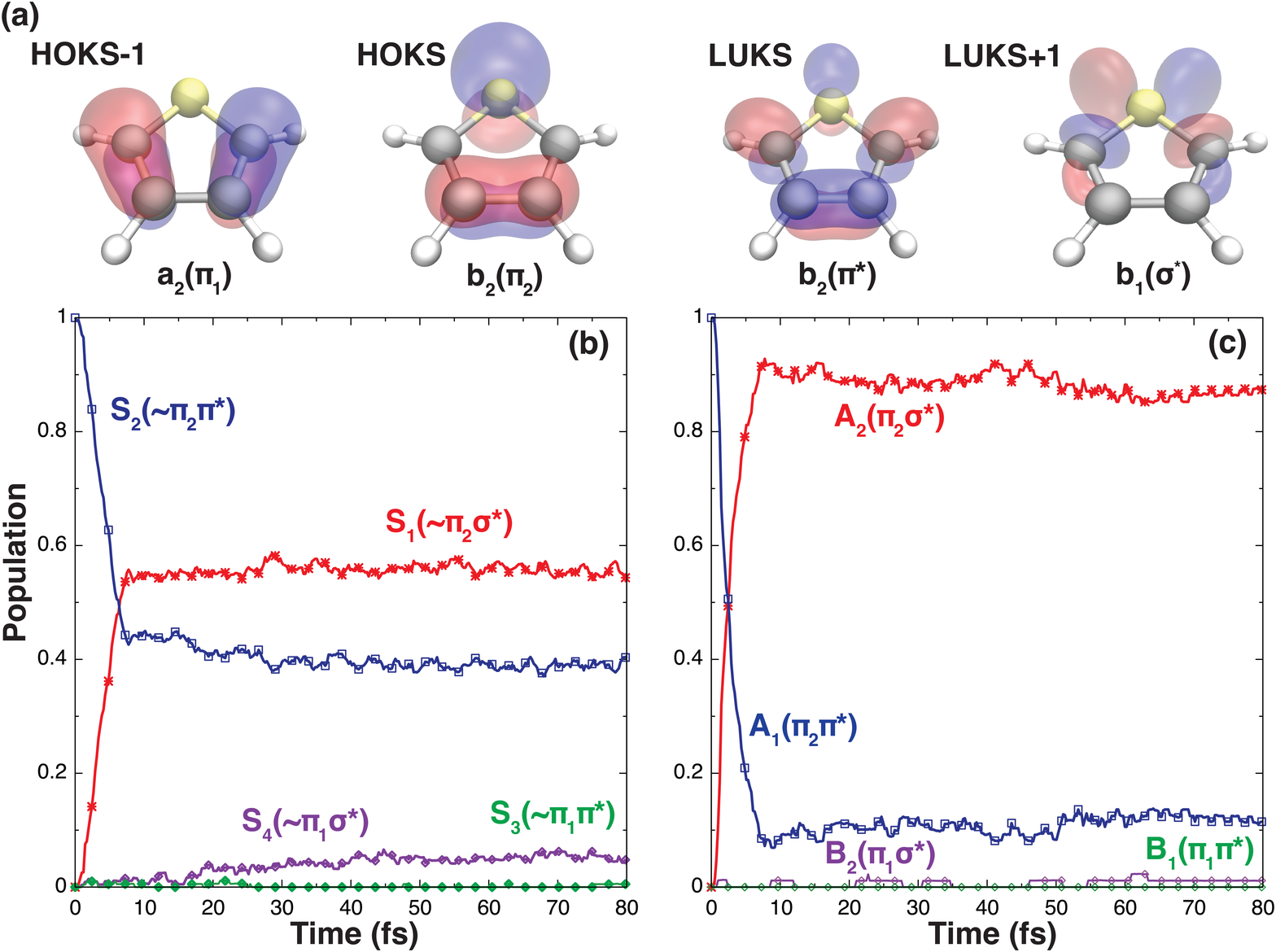}
  \caption{\textbf{(a)} The top contributing KS orbitals of the excited electronic states of selenophene referenced from the equilibrium geometry. \textbf{(b)}The SQC/MM population dynamics after sampling a $298$K harmonic oscillator Wigner distribution ($\approx 200$ trajectories). \textbf{(c)} The SQC/MM population dynamics with the positions initialized to the $C_{2v}$ geometry and the velocities sampled from a $298$K Boltzmann distribution ($\approx 100$ trajectories). 
}
  \label{Fig6}
\end{figure*}

The electronically nonadiabatic dynamics of selenophene were simulated by initially sampling $200$ nuclear positions and momenta directly from a $298$K ground-state harmonic oscillator Wigner distribution. The four lowest energy electronic states were included in the simulations which, as discussed below, have mixed $\pi\pi^*$-$\pi\sigma^*$ character due to out-of-plane distortions coupling together $\pi^*$ and $\sigma^*$ orbitals (shown in \ref{Fig6}(a)) in the Franck-Condon region. The electronic states ($S_1$, $S_2$, $S_3$, and $S_4$) were initially assigned to the $C_{2v}$ reference states ($A_1$, $A_2$, $B_1$, and $B_2$) described above according to their maximum overlap and these characters were monitored during the trajectories as defined by the multi-state tracking protocol. After Wigner sampling nuclear positions and momenta and assigning the corresponding electronic states, the electronic oscillator variables were initialized via the SQC protocol with the $S_2$ state initially populated, i.e. the state that overlapped most with the optically-bright $A_1(\pi_2\pi^*)$ state. The coupled nuclear and electronic DOF were allowed to propagate via Meyer-Miller dynamics on the potential energy surfaces with a $0.24$ fs time step for $80$ fs. As with malonaldehyde, the {$\gamma$}-adjustment protocol was employed in the initial SQC sampling protocol. 

The primary orbital contributions to the electronic transitions are shown in Fig.~\ref{Fig6}(a). The $a_2$ HOKS-1 orbital is a bonding $\pi$ orbital on the carbon backbone while the $b_2$ HOKS orbital is a combination of a bonding $\pi$-type orbital on the carbon backbone with a $\pi$ orbital on the selenium. In the valence space, the $b_2$ LUKS orbital has the same bonding $\pi$ structure as the HOKS orbital but is anti-bonding with the neighboring carbon atoms while the LUKS+1 orbital is a combination of anti-bonding $\sigma$-type orbitals on both the selenium atom and the carbon ring. The optically-bright $A_1(\pi_2\pi^*)$ and allowed, but dark, $B_1(\pi_1\pi^*)$ electronic states are an excitation from the HOKS and HOKS-1 orbitals to the LUKS orbital, respectively. Similarly, the optically-forbidden $A_2(\pi_2\sigma^*)$ and allowed, but dark $B_2(\pi\sigma^*)$ electronic states are an excitation from the HOKS and HOKS-1 orbitals to the LUKS+1 orbital, respectively.

As shown in Fig.~\ref{Fig6}, the population dynamics depend significantly on the character of the initially populated electronic state. When the initial geometries are Wigner sampled (see Fig. \ref{Fig6}(b)), the majority of population transfers between the $S_2$ and $S_1$ states before $t=10$ fs. After approximately $60$\% of the initial population has transferred into $S_1$, the exchange abruptly stops and the populations are maintained for the remainder of the dynamics---although some population (less than $10$\%) does transfer into the $S_4$ state. The $S_3$ electronic state doesn't acquire any substantial population on the time scales simulated. The ceasing of this abrupt exchange after $10$ fs is surprising, and as an additional experiment, designed to gauge the effect of exciting into a state of mixed character (as discussed above), $100$ trajectories were initialized with a single value of nuclear coordinates (precisely the $C_{2v}$ equilibrium geometry) with momenta sampled from a $298$K Boltzmann distribution. Though there does not appear to be an obvious justification for this, the idea was to explore the population dynamics that result from starting in the $A_1$ electronic state which has pure $\pi_2\pi^*$ character. As shown in Fig.~\ref{Fig6}(c), when the dynamics are initialized in this way more than $90$\% of the population transfers directly to the $A_2$ state and neither the $B_2$ or $B_1$ states acquire any substantial population.

\begin{figure}[t!]
\centering
  \includegraphics[height=11cm]{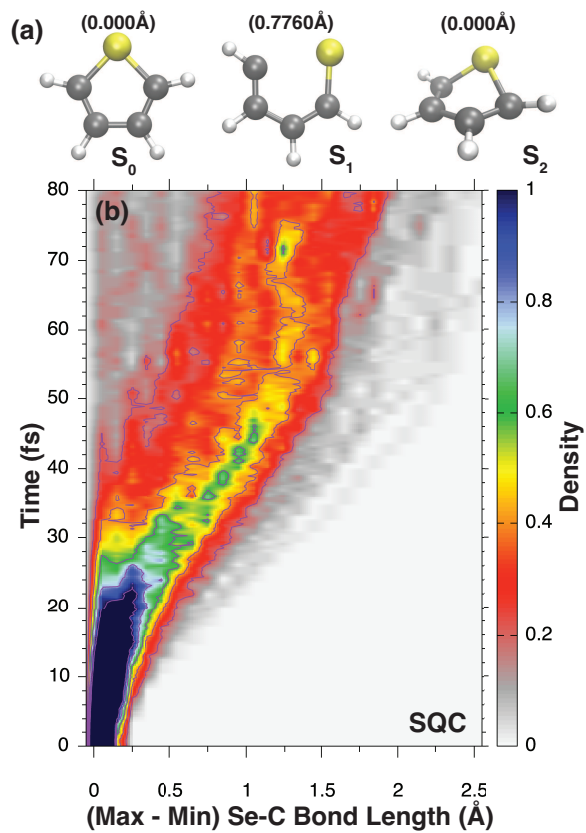}
  \caption{\textbf{(a)} Stationary points on the ground, first-, and second-excited electronic states of selenophene. \textbf{(b)} The difference between maximum and minimum selenium-carbon bond lengths monitored across the $\approx200$ Wigner-initialized SQC/MM trajectories. 
  }
  \label{Fig7}
\end{figure}

When referenced from the symmetric $C_{2v}$ geometry, vibronic predictions from the nonadiabatic dynamics follow a predictable trend in that $\pi\sigma^*$ states $(A_2, B_2)$ result in ring opened configurations and $\pi\pi^*$ states $(A_1, B_1)$ result in ring puckered configurations. However, when the initial state is mixed, as is the case after sampling the Wigner distribution, the vibronic pathways are mediated by the amount of $\pi\pi^*$ or $\pi\sigma^*$ character that is present on the effective potential energy surface. At stationary points on the adiabatic potential energy surfaces $S_0$ and $S_2$, the selenophene ring is closed as shown in Fig.~\ref{Fig7}(a). The stationary point on the $S_1$ potential energy surface is ring opened which corresponds to an optimized difference bond length, defined as the Max-Min bond lengths between selenium and the neighboring carbon atoms, of $0.776 {\AA}$. The ring opening dynamics are shown in Fig.~\ref{Fig7}(b) where this change in the Max-Min difference bond length was monitored and binned across the Wigner sampled trajectories, i.e. those corresponding to the population dynamics shown in Fig.~\ref{Fig6}(b). Clearly, by $t = 30$ fs most of the trajectories resulted in ring opening with most of trajectories after $t=50$ fs predicting difference bond lengths \emph{greater} than the optimized value on the $S_1$ potential energy surface. The majority of trajectories ring open ($\approx 85$\%) which is seemingly contradictory to the predicted population dynamics shown in Fig.~\ref{Fig6}(b) where only approximately $60$\% of the population transfers from the $S_2$ to the $S_1$ state. This evident contradiction can be understood as resulting from a substantial number of trajectories initially excited into $S_2$ having enough $\pi\sigma^*$ character such that their electronic configuration does not prevent ring opening. For comparison, $96$\% of the trajectories that were initialized to from the $C_{2v}$ equilibrium geometry, i.e. those corresponding to the population dynamics in Fig.~\ref{Fig6}(c), underwent ring opening.

\begin{table}[t!]
\centering
\renewcommand{\arraystretch}{1.4}
\caption{The trajectories that crossed the Coulson-Fischer point during the SQC/MM nonadiabatic dynamics simulations of selenophene.}
\begin{tabular}{lll}
Time ($t$,fs) & $\%$ traj. crossed C-F point  & $<S^2>$  \\
\hline
 0 & 0.00 & 0.000 \\
10 & 0.00 & 0.000 \\
20 & 0.00 & 0.000 \\
30 & 16.8 & 0.077 \\
40 & 54.1 & 0.312 \\
50 & 66.9 & 0.461 \\
60 & 75.6 & 0.604 \\
70 & 85.5 & 0.673 \\
80 & 88.4 & 0.743 \\
\end{tabular}
\label{tab2}
\end{table}

A potentially concerning aspect of the simulations of selenophene are the number of Wigner sampled trajectories that cross the Coulson-Fischer (C-F) point \cite{coulson1949xxxiv} as shown in Table \ref{tab2}. By $t = 30$ fs, {17\%} of the trajectories crossed the C-F point with nearly 88\% crossing by $t = 80$ fs. Since the trajectories were simulated using a restricted formalism, crossing the C-F point often results in an artificial increase of the potential energy as the Se-C ring is broken. Performing the simulations with an unrestricted KS determinant would seemingly correct for this issue as spin symmetry breaking would lower the potential energy as the ring is broken. However, unrestricted KS orbitals have been shown to result in nonphysical potential energy surfaces beyond the C-F point \cite{hait2019beyond}. In the event however that TDDFT/TDA is employed to simply predict whether ring opening will occur or not, crossing the C-F point during a dynamics trajectory is not too concerning since the C-F point is crossed on the $S_1$ potential energy surface which is repulsive along the bond-breaking coordinate.

\section{Conclusions}

The symmetric quasi-classical model for quantizing classical Meyer-Miller vibronic dynamics is an efficient, and often quite accurate framework for performing \textit{ab inito} molecular dynamics for electronically nonadiabatic processes, such as vibrational-DOF enhanced electronic energy transfer dynamics and the role that nonadiabatic energy transfer has on geometric and other properties. Here, what has been developed for general use is an implementation of the SQC/MM model using ``on-the-fly'' TDDFT/TDA within the widely available Q-Chem quantum chemistry software package, including the efficient implementation of new algorithms that improve the compute cost when evaluating analytic nuclear gradients and first-order derivative coupling vectors. In particular, new digestion routines were proposed that contract the full set of density matrices with a common set of integrals and integral derivatives which were shown to speedup the calculations by a factor of four compared with the brute force method. The efficiency gains that were achieved as a result of these new algorithms should aid in simulating realistic time-scales of nonadiabatic dynamics in moderately-sized molecular systems. 

As an illustrative example of this new implementation, the excited-state hydrogen transfer dynamics of malonaldehyde were analyzed. The simulations presented here suggest that when malonaldehyde is photoexcited to the $S_2$ state, nonradiative decay occurs rapidly to the nearby $S_1$ state where more than $50$\% of the population is transferred before $t = 50$ fs. In the intermediate regime ($t<50$ fs), the hydrogen atom, which bonds together the ring structure, shuffles rapidly back and forth between the donor and acceptor oxygen terminals. Once sufficient population has transferred into the $S_1$ state however, the hydrogen atom mainly localizes on either of the two oxygen terminals. These simulations present a computationally simple example of the accuracy of TDDFT/TDA in combination with the SQC/MM approach when compared with other comparable nonadiabatic dynamics methods. 

The ring-opening dynamics of selenophene were also investigated which posed some challenges for TDDFT/TDA due to the Se-C bond breaking \emph{after} crossing the C-F point. The simulations predict that after photoexcitation to the $S_2$ state population transfers very rapidly to the $S_1$ state with more than $60$\% transferring before $t = 10$ fs. After approximately $20$ fs, either by sufficient population accruing in the $S_1$ state or the initialized $S_2$ state having sufficient $\pi\sigma^*$ character, the Se-C bond breaks resulting in ring opening. Making vibronic predictions, i.e identifying specific electronic rearrangements and configurations that are directly responsible for ring opening, was challenging in the case of selenophene as a result of mixing between the $\pi^*$ and $\sigma^*$ orbitals near the Franck-Condon region. 

A serious limitation when using TDDFT/TDA with nonadiabatic dynamics methods is the incorrect topology predictions of conical intersections between ground and excited electronic states. While the malonaldehyde and selenophene simulations presented here predicted the population dynamics between excited electronic states only, these systems are known to undergo nonradiative decay to the ground state which is a physically relevant pathway that was neglected. Some electronic structure approaches, such as spin-flip variants of TDDFT/TDA \cite{zhang2014analytic,zhang2021nonadiabatic,harabuchi2013automated,yue2018performance,salazar2020theoretical}, have been developed already that address the challenges when calculating first-order derivative coupling vectors between ground and excited electronic states. Efficiently implementing these approaches in the framework of SQC/MM will be the result of future work.

\section{Conflicts of interest}
There are no conflicts to declare.

\section{Acknowledgements}
The authors thank Bill Miller for support and encouragement and without whom this work would certainly not be possible. This work is supported by the Director, Office of Science, Office of Basic Energy Sciences of the US Department of Energy under contract No. DE-AC02-05CH11231. This work is supported by the National Science Foundation under grant number CHE-1856707. This research used computational resources of the National Energy Research Scientific Computing Center, a DOE Office of Science User Facility supported by the Office of Science of the U.S. Department of Energy under Contract No. DE-AC02-05CH11231.

\begin{appendix}
\section{Density Matrix Derivations}

The matrix elements of \textbf{A} are the response of the KS Fock matrix ($\textbf{F}$) to a perturbation in the one-particle density matrix ($\textbf{P}$) \cite{liu2010parallel,hirata1999time}. The matrix elements are
\begin{equation*}
    A_{ai,bj} = \frac{\partial F_{ai}}{\partial P^{bj}} = (\epsilon_{a} - \epsilon_{i})\delta_{ab}\delta_{ij} + \Pi_{ai,bj} + \Omega_{ai,bj},
\end{equation*}
\noindent which include the energies of the occupied and virtual KS orbitals, the two-electron integral tensor ($\boldsymbol{\Pi}$), with elements $\Pi_{ai,bj} = (ia|jb) - C_{HF}(ij|ab)$ where $C_{HF}$ is a scalar denoting the percent Hartree-Fock exchange, and 
\begin{equation*}
    \Omega_{ai,bj} = \frac{\partial F_{xc,ai}}{\partial P^{bj}},
\end{equation*}
\noindent which is the response of the exchange-correlation Fock Matrix to a perturbation in the one-particle density matrix. The exchange-correlation Fock matrix is the response of the exchange correlation energy ($E_{xc}$) to the same perturbation
\begin{equation*}
    F_{xc,ai} = \frac{\partial E_{xc}}{\partial P^{ai}} = \int \sum_\xi \frac{\partial f_{xc}}{\partial \xi} \frac{\partial \xi}{\partial P^{ai}}d\textbf{r}, 
\end{equation*}
\noindent where $\{\xi\}$ denotes a set of independent parameters defined in the exchange-correlation functional ($f_{xc}$) and which depend linearly on the one-particle density matrix. 

The differentiation of these matrix elements are typically performed in the AO basis and contracting the derivatives with relaxed (denoted by ') one- and two-particle density matrices is required when building the nonadiabatic coupling vector \textbf{h}$_{IJ}$ \cite{zhang2014analytic,ou2015derivative}. The ground to excited state one-particle transition density matrix is
\begin{equation}\label{e:Td}
    \textbf{T}^I = \textbf{C}_v\textbf{X}^{I}\textbf{C}_o^{\dag},
\end{equation}
\noindent where $\textbf{C}_o$ and $\textbf{C}_v$ are rectangular matrices that contain the occupied and virtual blocks of the KS orbital coefficient matrix $\textbf{C}$. The generalized difference density matrix (i.e. when $I=J$, the ground to excited state difference density matrix is obtained) is
\begin{equation*}
    \textbf{P}^{IJ}_{\Delta} = \frac{1}{2} \textbf{C}_v(\textbf{X}^{I}\textbf{X}^{J\dag} + \textbf{X}^{J}\textbf{X}^{I \dag})\textbf{C}_v^{\dag} \\ - \frac{1}{2}\textbf{C}_o(\textbf{X}^{I\dag}\textbf{X}^{J} + \textbf{X}^{J\dag}\textbf{X}^{I})\textbf{C}_o^{\dag},
\end{equation*}
\noindent which depends explicitly on the occupied-occupied and virtual-virtual blocks of KS orbital coefficient matrix. The relaxed generalized difference density matrix
\begin{align}\label{e:Pdrel}
    \textbf{P}^{IJ'}_{\Delta} = \textbf{P}^{IJ}_{\Delta} + \textbf{P}^{IJ}_{Z} = \textbf{P}^{IJ}_{\Delta} +    \textbf{C}_v\textbf{Z}^{IJ}\textbf{C}_o^{\dag} + \textbf{C}_o\textbf{Z}^{IJ\dag}\textbf{C}_v^{\dag},
\end{align}
\noindent is obtained after differentiating the KS orbital coefficients. This, in turn, requires solving the coupled-perturbed self-consistent field (CPSCF) equations for a Z-vector ($\textbf{Z}^{IJ})$ between states $I$ and $J$
\begin{equation}\label{e:cpscf}
    \bigg(\textbf{E}_{KS}^{\Theta\Theta}\bigg)\textbf{Z}^{IJ} = \textbf{L}^{IJ}
\end{equation}
\noindent where $\Theta$ denotes the set of virtual-occupied orbital rotations and $\textbf{L}^{IJ}$ is a Lagrangian. The components of the CPSCF equations are defined as
\begin{equation*}
    \bigg(\textbf{E}_{KS}^{\Theta\Theta}\bigg)\textbf{Z}^{IJ} \equiv \textbf{C}_v^{\dag}\textbf{F}\textbf{C}_v\textbf{Z}^{IJ} - \textbf{Z}^{IJ}\textbf{C}_o^{\dag}\textbf{F}\textbf{C}_o \\ + \textbf{C}_v^{\dag}\bigg[(\boldsymbol{\Pi} + \boldsymbol{\Omega}) \cdot \textbf{P}_{Z}^{IJ} \bigg]\textbf{C}_o
\end{equation*}

\noindent where $\textbf{P}_Z^{IJ}$ is given in \Eq{Pdrel} and
\begin{multline*}\label{e:lag}
    \begin{split}
     \textbf{L}^{IJ} &\equiv \textbf{C}_v^{\dag}\bigg[(\boldsymbol{\Pi} + \boldsymbol{\Omega})\cdot \textbf{P}^{IJ}_{\Delta} +\textbf{T}^{I \dag} \cdot \boldsymbol{\Xi} \cdot \textbf{T}^{J} \bigg]\textbf{C}_o - \\
     &\frac{1}{2}\textbf{C}_v^{\dag}\bigg[(\boldsymbol{\Pi} + \boldsymbol{\Omega})\cdot \textbf{T}^{I \dag}\bigg]\textbf{C}_v\textbf{X}^{J} - \frac{1}{2}\textbf{C}_v^{\dag}\bigg[(\boldsymbol{\Pi} + \boldsymbol{\Omega})\cdot \textbf{T}^{J \dag}\bigg]\textbf{C}_v\textbf{X}^{I} + \\
     &\frac{1}{2}\textbf{X}^{J}\textbf{C}_o^{\dag}\bigg[(\boldsymbol{\Pi} + \boldsymbol{\Omega})\cdot \textbf{T}^{I \dag}\bigg]\textbf{C}_o - \frac{1}{2}\textbf{X}^{I}\textbf{C}_o^{\dag}\bigg[(\boldsymbol{\Pi} + \boldsymbol{\Omega})\cdot \textbf{T}^{J \dag}\bigg]\textbf{C}_o,
     \end{split}
\end{multline*}
\noindent which defines the Lagrangian. The solution to \Eq{cpscf} requires contracting the two-electron integrals and second-functional derivatives with the generalized difference and transition density matrices. Additionally, the third functional derivative of the exchange-correlation energy 
\begin{equation*}
    \boldsymbol{\Xi}_{\mu\nu,\lambda\sigma,\kappa\gamma} = \frac{\partial \Omega_{\mu\nu,\lambda\sigma}}{\partial P^{\kappa\gamma}},
\end{equation*}
\noindent is contracted with the transition density matrices of states $I$ and $J$. With the corresponding Z-vector, the relaxed generalized difference density is constructed according to \Eq{Pdrel} and this matrix is contracted with the core Hamiltonian ($\textbf{H}^{[\textbf{R}]}$) and exchange-correlation Fock ($\boldsymbol{F_{xc}^{[\textbf{R}]}}$) integral derivatives when building the nonadiabatic coupling vector.

Additionally, evaluation of the nonadiabatic coupling vector in \Eq{hij} requires contracting the two-particle ($\boldsymbol{\Gamma}^{IJ'}$) and energy-weighted ($\boldsymbol{W}^{IJ'}$) density matrices with the two-electron ($\boldsymbol{\Pi}^{[\textbf{R}]}$) and overlap ($\textbf{S}^{[\textbf{R}]}$) integral derivatives, respectively. These matrices are defined accordingly as
\begin{subequations} \label{e:tpdm}
   \begin{equation}
    \Gamma^{IJ'} = \bigg(\textbf{P} \otimes \textbf{P}_{\Delta}^{IJ'}\bigg) + \bigg(\textbf{T}^{I\dag} \otimes \textbf{T}^{J}\bigg)
    \end{equation} 
    \begin{equation}
    \textbf{W}^{IJ'} = -\frac{1}{2}\boldsymbol{\Lambda}^{IJ'}\textbf{C}\textbf{C}^{\dag} - \frac{1}{2}\textbf{C}\textbf{C}^{\dag} \boldsymbol{\Lambda}^{IJ'\dag} 
    \end{equation}
\end{subequations}
\noindent where
    \begin{equation*}
   \begin{split}
        \boldsymbol{\Lambda}^{IJ'} = \textbf{P}_{\Delta}^{IJ'} \cdot \textbf{F} + \textbf{P}\bigg[(\boldsymbol{\Pi} + \boldsymbol{\Omega}) \cdot \textbf{P}_{\Delta}^{IJ'}  + \textbf{T}^{I\dag}\cdot \boldsymbol{\Xi}\cdot\textbf{T}^{J}\bigg] \\
        + \frac{1}{2} \textbf{T}^{I}\bigg[(\boldsymbol{\Pi}+\boldsymbol{\Omega})\cdot\textbf{T}^{J\dag}\bigg] + \frac{1}{2}\textbf{T}^{J}\bigg[(\boldsymbol{\Pi}+\boldsymbol{\Omega})\cdot\textbf{T}^{I\dag}\bigg] \\
        +\frac{1}{2} \textbf{T}^{I\dag}\bigg[(\boldsymbol{\Pi}+\boldsymbol{\Omega})\cdot\textbf{T}^{J} \bigg] + \frac{1}{2}\textbf{T}^{J\dag}\bigg[(\boldsymbol{\Pi}+\boldsymbol{\Omega})\cdot\textbf{T}^{I}\bigg],
    \end{split}
   \end{equation*}

\noindent which are notably relaxed due to their dependence on the relaxed generalized difference density matrix from \Eq{Pdrel}. 

\section{Timing Analysis}

\begin{figure}[t!]
  \centering
  \includegraphics[height=12cm]{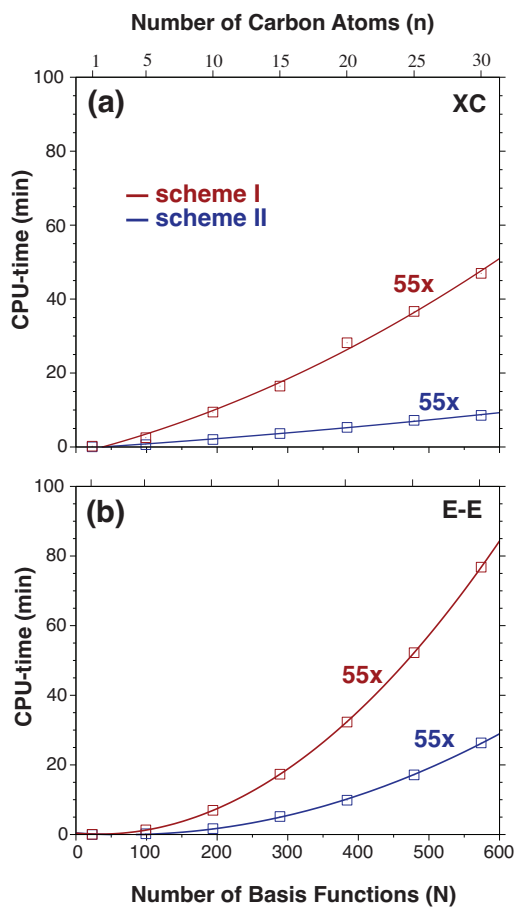}
  \caption{Timing data as a function of basis set size and increasing chain length for the the exchange-correlation (XC) integral derivative contractions \textbf{(a)} and the electron-electron (E-E) repulsion integral derivative contractions \textbf{(b)}. The red curves (scheme I) and the blue curves (scheme II) correspond to a polynomial least-squares fit of the timings of each component.}
  \label{FigA1}
\end{figure}

The timings were analyzed by fitting the data to a simple quadratic polynomial
\begin{equation*}
    \text{CPU-time}(N) = aN^2 + bN + c,
\end{equation*}
\noindent where $N$ denotes the total number of basis functions. The fit parameters of the most computationally expensive components of the nonadiabatic coupling are provided in Table \ref{tab1}. Both the exchange-correlation (XC) and electron-electron (E-E) components of the nonadiabatic coupling vector scale quadratically with the number of basis functions. However, the quadratic prefactor corresponding to the exchange-correlation term, after employing scheme II, is reduced by a factor of $\approx 10$ compared with scheme I. The leading prefactor for the electron-electron integral derivative contractions, which is clearly the most computationally expensive step, is reduced by $\approx2.5$ after employing scheme II.
\begin{table}[t]
\centering
\renewcommand{\arraystretch}{1.4}
\caption{Fit parameters for the exchange-correlation (XC) integral derivative contraction (solid lines in Fig.~\ref{FigA1}(a)) and the electron-repulsion (E-E) integral derivative contraction (solid lines in Fig.~\ref{FigA1}(b)) components of the analytic nuclear gradient and nonadiabatic coupling vector (corresponding to Fig. \ref{FigA1}) with increasing alkane chain length.}
\begin{tabular}{llll}
Component ($\textbf{h}_{IJ})$ &$\phantom{-}a\times10^{5}$&$\phantom{-}b\times10^{2}$&$\phantom{-}c\times10^{1}$\\
\hline
XC (scheme I) & $\phantom{-}6.769$ & $ \phantom{-}4.475$ & $-18.830$\\
XC (scheme II) & $\phantom{-}0.661$ & $ \phantom{-}1.224$ & $-4.489$\\
\hline 
E-E (scheme I) & $\phantom{-}26.14$ & $-1.696$ & $\phantom{-}3.773$\\
E-E (scheme II) & $\phantom{-}10.23$ & $-1.389$ & $\phantom{-}4.397$\\
\end{tabular}
\label{tab1}
\end{table}
Timing illustrations for the speedups possible when employing scheme~II are shown for the the exchange-correlation (\textbf{XC}, Fig.~\ref{FigA1}(a)) and electron-electron (\textbf{E-E}, Fig.~\ref{FigA1}(b)) integral derivative contractions. Fitting the timings with a quadratic polynomial reveals that a factor of $\approx 3$ speedup for the electron-electron repulsion integral derivative contractions is possible. The exchange-correlation integral derivative contractions result in a $\approx 10\times$ speedup.

\end{appendix}

\clearpage
\printbibliography
\end{document}